# Graphene on quartz modified with rhenium oxide as a semitransparent electrode for organic electronic

Paweł Krukowski[a*], Michał Piskorski[a], Ruslana Udovytska[b], Dorota A. Kowalczyk[a], Iaroslav Lutsyk[a], Maciej Rogala[a], Paweł Dąbrowski[a], Witold Kozłowski[a], Beata Łuszczyńska[b], Jarosław Jung[b], Jacek Ulański[b], Krzysztof Matuszek[b], Aleksandra Nadolska[a], Przemysław Przybysz[a], Wojciech Ryś[a], Klaudia Toczek[a], Rafał Dunal[a], Patryk Krempiński[a], Justyna Czerwińska[a], Maxime Le Ster[a], Marcin Skulimowski[c], Paweł J. Kowalczyk[a]

[a] Department of Solid State Physics (member of National Photovoltaic Laboratory, Poland), Faculty of Physics and Applied Informatics, University of Lodz, Pomorska 149/152, 90–236 Łódź, Poland

[b] Department of Molecular Physics (member of National Photovoltaic Laboratory, Poland), Lodz University of Technology, Żeromskiego 116, 90–924 Łódź, Poland

[c] Department of Intelligent Systems, Faculty of Physics and Applied Informatics, University of Lodz, Pomorska 149/152, 90–236 Łódź, Poland

## Abstract

Our research shows that commercially available graphene on quartz modified with rhenium oxide meets the requirements for its use as a conductive and transparent anode in optoelectronic devices. The cluster growth of rhenium oxide enables an increase in the work function of graphene by 1.3 eV up to 5.2 eV, which guarantees an appropriate adjustment to the energy levels of the organic semiconductors used in OLED devices.

## 1. Introduction

The displays of many common devices such as TV sets, mobile phones and cameras are built on organic light-emitting diodes (OLEDs) due to unrivalled image quality resulting from accurate color reproduction. The amazing and very rapid progress in the development of the OLEDs manufacturing technology makes it widely regarded as the technology of the future [1]. One of the most important elements in the construction of an OLED is a highly conductive and transparent electrode. Currently, in typical OLEDs design an electrode is made of an indium tin oxide (ITO) layer deposited on glass. Despite its many advantages, the ITO layer has the very serious disadvantage of brittleness. This makes construction of flexible displays with ITO layer highly challenging. Therefore, at present, there is an intense search for alternative materials that can be used as flexible electrodes in OLEDs and photovoltaic cells [2].

It is well known that graphene conducts electricity and heat very well, is transparent in the visible range of electromagnetic radiation, and is also highly flexible. All the above characteristics make it a great candidate as a potential conductive and transparent electrode in optoelectronic devices such as OLEDs and photovoltaic cells. However, graphene in its undoped form is characterized by a too low work function (3.9 – 4.5 eV), which makes it impossible to adjust the energy levels of organic semiconductors used to make the optoelectronic devices [3,4]. One of the methods of adjusting the energy levels of graphene is through applying a layer of transition metal oxide [5–7], resulting in a hybrid system with the desired properties. An excellent example of this approach is the work of Meyer *et al*. [6], in which they showed an OLED with an anode based on a trilayer graphene deposited on glass, which was modified with a thin (5-nm) layer of molybdenum oxide ($MoO_3$). We have recently showed that the same effect can be obtained by using a thinner layer of crystalline molybdenum oxide [8–11]. We have also made attempts to modify the work function of graphene using other transition metal oxides such as rhenium oxide [5] and titanium oxide [12]. Our original approach to obtain a transition metal oxide layer is first to optimize the process of fabricating the appropriate layer on the surface of highly oriented pyrolytic graphite (HOPG), which is similar to graphene. In the next step, an analogous growth is carried out on graphene as shown in the works [8,9]. During the research, we were able to fabricate a fully functional OLED with a graphene electrode. For this purpose, we were the first in the world to use a bilayer graphene, which was epitaxially fabricated on the surface of a silicon carbide crystal SiC(0001) using the chemical vapor deposition (CVD) method (ITME). Rhenium oxide was used to increase the work function of the bilayer graphene [5]. We demonstrated that such a hybrid system worked in OLEDs as a conductive and transparent anode. However, due to the very high cost of bilayer graphene on SiC, we were not able to fully optimize the structure and fabrication processes of OLEDs. Therefore, we decided to use in further research about 10 times cheaper equivalent –graphene on quartz, which is commercially offered by MSE Supplies LLC (USA), in which graphene is prepared on copper foil by CVD method and then transferred and then transferred onto quartz using polymer: poly(methylmethacrylate) (PMMA) in wet transfer process. The manufacturer guarantees the sheet resistance of the tested graphene on quartz at a very low level of only 360 ± 30 Ω/□, this value is lower than the sheet resistance of graphene on SiC (> 600 Ω/□) [13].

It is worth noting that 2D materials, such as graphene and graphene oxide can be used in optoelectronic devices not only as a conductive electrode, but also as a hole injection, blocking or transport layer. For example, graphene oxide on ITO can be used as a hole injection layer in OLEDs which leads to a reduction of the device threshold voltage to a value of 3.0 V, i.e., a lower value than the commonly used polymeric layer of a mixture of two PEDOT:PSS ionomers (3.3 V) [14].

In this paper, we present the results of our study on the modification of the graphene work function with the use of thermally deposited rhenium oxide under ultrahigh vacuum conditions (pressure ~ $6 \times 10^{-9}$ mbar). Our studies show that relatively cheap graphene on quartz modified with rhenium oxide meets all the basic requirements that are set for a transparent anode and can be successfully used in place of the much more expensive bilayer graphene on SiC(0001).

## 2. Results

Figure 1(a) shows the topographic image of the graphene on quartz obtained using an atomic force microscope (AFM). The Ntegra Aura AFM microscope (NT-MTD) operating in semi-contact mode under ambient conditions was used. Our AFM study shows that the graphene layer coverage is better than 95%, it is continuous with marginal defects and polymer contaminations. The graphene is covered by small amount of contaminations (PMMA polymer residues) observed as bright features, which persisted on graphene after dissolution of polymer layer acting as the supporting layer in the wet transfer process [15,16]. Additionally, some typical graphene defects such as dark cracks and bright wrinkles can be seen in Fig. 1(a). The presence of wrinkles and cracks can be explained in terms of the mechanical handling of graphene during transfer to the substrate and presence of wrinkles in the as-grown graphene on Cu foil [16]. The crystallinity and roughness of Cu substrate is one of the most important factor that influence the formation of wrinkles and cracks in as-grown graphene during cooling in CVD process [17]. Note that a minimum critical length is necessary for wrinkling of graphene layer and stabilizing the wrinkles. Graphene wrinkles with lengths smaller than a critical value are not stable and can be removed by thermal annealing [18]. Atomic defects existing on graphene film were not seen using AFM due to limited resolution. Figure 1(b) shows an AFM image of rhenium oxide, with a thickness of 10 nm, deposited on graphene by thermal deposition from a crucible filled with $Re_2O_7$ powder under ultrahigh vacuum. The thickness of the rhenium oxide layer was estimated using a quartz microbalance. Here the thickness refers to hypothetical uniform single-crystalline multilayer of $Re_2O_7$ grown on substrate. Our AFM study shows that the growth of compound with the formula $Re_2O_7$ formula on graphene is cluster-like. This indicates that the diffusion mean free path of arriving compounds is very short due to possible existence of atomic defects in graphene structure. Presumably, these defects are working as pin sites immobilizing diffusing rhenium oxide clusters. This combined with the relatively strong rhenium oxide / rhenium oxide interaction, leads to the growth of separated clusters according with the Volmer-Weber (VW) model [19]. This explains why atomically flat crystalline layers of rhenium oxide on graphene could not be obtained in our study despite great experimental efforts. Moreover, for small coverages we observed high mobility of rhenium oxide clusters on graphene, which made correct AFM imaging difficult, but which supports the hypothesis of the VW model. It also shows that the bonding of the clusters to the pin sites is relatively week allowing for their manipulation using AFM. Our study shows that the growth of rhenium oxide on graphene deposited on quartz is similar to the growth of rhenium oxide on graphene obtained on SiC(0001) [5]. To qualitatively analyze the chemical composition of the clusters formed on graphene, X-ray photoelectron spectroscopy (XPS) studies were performed. Figure 2(a) shows the XPS results in the electron binding energy region of the core 4f states for rhenium with a thickness corresponding to a 10 nm layer. It should be noted that we annealed the graphene at 325 °C to remove PMMA polymer residues before depositing rhenium oxide. A Phoibos 150 hemispherical electron energy analyzer (SPECS) with a 2D-CCD detector was used in this study. The sample was exposed to radiation from a DAR 400 lamp operating with a Mg Kα electrode at an energy of 1253.64 eV. CasaXPS software was used to perform the deconvolution of the individual spectral lines. The analysis of the spectra began by removing the background using the Shirley method. A Gauss-Lorentz product function fit (GL(75)) was then performed with a full width at half maximum (FWHM) for all lines of 1.7 eV. Due to the spin-orbit coupling, the occurrence of the characteristic $4f_{7/2}$–$4f_{5/2}$ doublet with an energy splitting of 2.4 eV with an integral intensity of 4:3 was considered in the analysis of the Re 4f spectrum. The XPS spectra show the occurrence of two different rhenium oxide phases. Mainly, the $Re_2O_7$ phase is observed with the characteristic $4f_{7/2}$–$4f_{5/2}$ doublet with energies of 45.3 eV and 47.7 eV, respectively, which is comparable with literature data [20]. In addition, we observed a $ReO_3$ phase with a characteristic $4f_{7/2}$–$4f_{5/2}$ doublet in small amounts with energies of 42.9 eV and 45.3 eV, respectively. Since $ReO_3$ exhibit a metallic conductivity similar to that of copper, we expect local conductivity enhancement. Similar results were also obtained for rhenium oxide with different thicknesses, which was deposited onto graphene on SiC(0001) by thermal deposition from a crucible filled with $ReO_3$ powder under ultrahigh vacuum conditions, where we also observed an increase in the $Re_2O_7$ phase [5]. To determine the work function of graphene modified with rhenium oxide, we performed ultraviolet photoelectron spectroscopy (UPS) studies. Figure 2(b) shows UPS studies for pure graphene annealed at 325 °C and for annealed graphene modified with rhenium oxide with a thickness corresponding to a 10 nm layer. For the study, we used a HeIα line of energy 21.23 eV using a HIS-13 UV lamp. To correctly measure the backscattered electron emission edge, a negative potential of 3.1 V was applied to the sample. Calibration of the UPS results was done by determining the Fermi level for the pure Au(111) surface. The black curve in Fig. 2(b) shows the UPS results collected for pure graphene. The work function estimated by determining the backscattered electron emission edge for pure graphene on quartz is 3.9 eV. This value is smaller than the work function obtained in our previous study for bilayer graphene on SiC(0001), which was 4.19 eV [5]. On the other hand, the red curve in Fig. 2(b) shows the UPS results for graphene modified with rhenium oxide of a thickness corresponding to a 10 nm layer. The work function estimated for the modified graphene is 5.2 eV. The observed very large increase in the value of work function for the modified graphene by +1.3 eV can be attributed to the proximity effect at the $Re_2O_7$/graphene interface. In addition, states originating from O 2p can be observed in the valence band at energies around 5 eV and states (O-Re) which are a mixture of rhenium/oxygen states at energies around 8 eV below the Fermi level. Certainly, the observed work function for graphene on quartz modified with a rhenium oxide confirms the possibility of using such a system as a transparent, conducting anode in OLEDs and photovoltaic cells. Analogous result of the increase in the graphene work function was obtained for $Re_2O_7$, which was deposited on a bilayer graphene on SiC(0001) [5].

In the final step, we performed a study using Raman spectroscopy. This study was designed to evaluate the continuity of the graphene layer on quartz and to determine whether $Re_2O_7$ thermally applied to the graphene surface causes defects in its structure. A Raman spectrometer (SOL Instruments) equipped with a 600 gr/mm diffraction grating and a 532 nm continuous-wave green laser (Coherent) at 5 mW with a 50× magnification objective (Zeiss) was used for the study. Figure 3 shows the Raman spectroscopy results of graphene on quartz after thermal deposition of $Re_2O_7$ with a thickness corresponding to a 50 nm layer. The spatial mapping mode with a size of 20 x 20 $\mu m^2$ was applied. The spectrum averaged from all 2500 spectra collected during the mapping is shown on the left. The following images show successively spatial maps of the intensity distribution for D-band (~1345 $cm^{-1}$), G-band (~1588 $cm^{-1}$) and 2D-band (~2685 $cm^{-1}$). The map for D-band shows the defect distribution of graphene after $Re_2O_7$ deposition. Raman spectroscopy studies clearly show (i) the presence of only local graphene defects, (ii) thermal deposition of rhenium oxide does not defect graphene, and (iii) the continuity of the layer is unaffected. Based on these observations, we believe that such a layer is suitable for use as a transparent conducting electrode in optoelectronic devices.

## 3. Conclusions

In conclusion, our study shows that commercially available graphene transferred from copper foil to quartz, further modified with rhenium oxide meets the basic requirements for use as a conductive and transparent anode in optoelectronic devices. The results of AFM studies show that the thermal growth of rhenium oxide on graphene on quartz is cluster-like according to the VW model. XPS studies have determined that the sample has a predominantly $Re_2O_7$ phase. UPS studies indicate that the work function of graphene can be easily modified by depositing $Re_2O_7$. We observed a very large increase in the work function for the modified graphene by as much as +1.3 eV to a level of 5.2 eV, that is acceptable for use in real-world devices. Raman spectroscopy studies indicate only local occurrence of defects after $Re_2O_7$ deposition. Intensive work is currently underway to use graphene deposited on quartz modified with $Re_2O_7$ as an anode in OLED design.

# Authors' statement

# Acknowledgements

This work was financially supported by the National Science Centre (Poland) under grants 2016/21/B/ST5/00984 (P.K., M.P., R.U., J.J., J.U.) and 2020/37/B/ST5/03929 (M.R., W.K., B.Ł., P.J.K.).

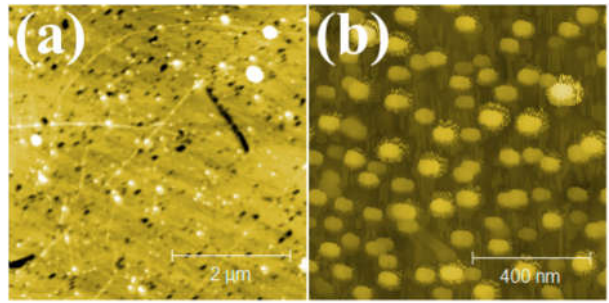

**Fig.1.** AFM topographic images of (a) graphene on quartz transferred from copper, (b) rhenium oxide with a thickness corresponding to a 10 nm layer deposited on the surface of a graphene/quartz system.

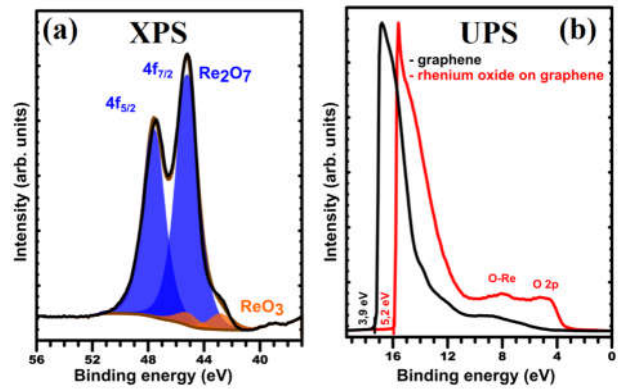

**Fig.2.** (a) XPS studies of graphene modified with rhenium oxide of a thickness corresponding to a 10 nm layer in the electron binding energy region of the core 4f states for rhenium. (b) UPS studies over a wide energy range for pure graphene and for modified graphene with rhenium oxide of a thickness corresponding to a 10 nm layer.

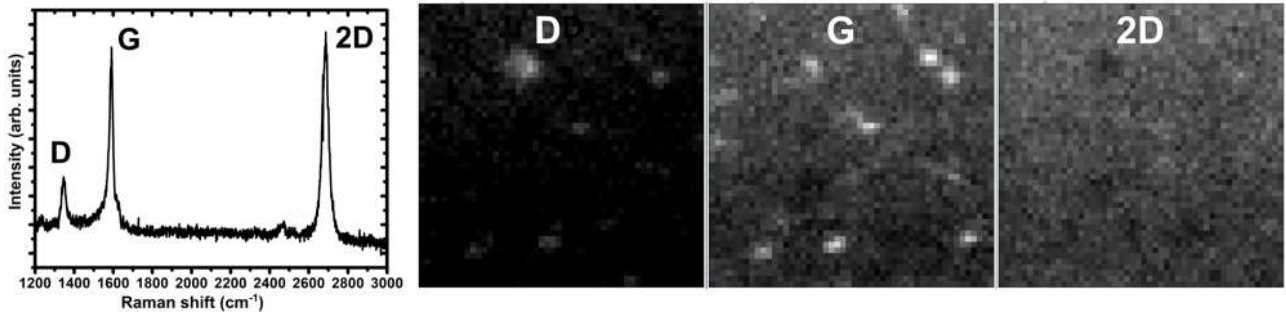

**Fig.3.** Investigation of graphene on quartz after thermal application of Re$_2$O$_7$ with a thickness corresponding to a 50 nm layer using Raman spectroscopy in spatial surface mapping mode. A 20 × 20 μm$^2$ Raman map was collected with a resolution of 50 × 50 measurement points. On the left, the spectrum averaged from all 2500 spectra collected during the mapping is shown. Next, the spatial intensity distribution maps for the D, G, and 2D bands can be seen sequentially.